\documentclass[aps,prb,twocolumn]{revtex4-1}

\usepackage{graphicx}

\begin{document}

\title{vdW-DF study of energetic, structural, and\\
vibrational properties of small water clusters and ice $I_h$}

\author{Brian Kolb}
\affiliation{Department of Physics, Wake Forest University, Winston-Salem, NC, 27109}
\author{T. Thonhauser}
\affiliation{Department of Physics, Wake Forest University, Winston-Salem, NC, 27109}
\date{\today}

\begin{abstract}
We present results for a density functional theory study of small
water clusters and hexagonal ice $I_h$, using the van der Waals
density functional (vdW-DF). In particular, we examine energetic,
structural, and vibrational properties of these systems. Our results exhibit
excellent agreement with both experiment and quantum-chemistry
calculations and show a systematic and consistent improvement over
standard exchange-correlation functionals---making vdW-DF a promising
candidate for resolving longstanding difficulties with density
functional theory in describing water. In addition, by comparing our
vdW-DF results with quantum-chemistry calculations and other standard
exchange-correlation functionals, we shed light on the question of why
standard functionals often fail to describe these systems accurately.
\end{abstract}

\maketitle

\section{Introduction}

\noindent
Despite the overwhelming abundance of water on Earth's surface and its
composing the majority of all known life, much of the physics of water
remains a mystery.  Unraveling that mystery requires the ability to
take microscopic information and extend it toward the macroscopic
regime.  Classical force fields and other parameterized models can
calculate average macroscopic properties of water
well,\cite{Guillot2002,Abascal2009,Sanz2004} but generally yield
little insight into the microscopic physics underpinning these
properties. On the other hand, quantum-chemistry methods such as
M{\o}ller-Plesset perturbation theory and coupled-cluster techniques
have been invaluable in understanding numerous molecular systems and
indeed are responsible for much of what is understood about water
today.\cite{Xantheas1993, Xantheas2002, Gregory1997, Shields2010} But,
their poor scaling with system size limits their use to systems
containing only a few molecules.\cite{Head-Gordon2008} Density
functional theory (DFT) scales substantially better and is thus, in
principle, well suited to make the connection to the macroscopic
regime.  Unfortunately, standard functionals fail to adequately
describe the weak van der Waals interactions that are critical to
obtain accurate diffusive and structural properties of
water.\cite{schwegler:5400, fernandez-serra:11136, lee:154507,
  guidon:214104, Sit2005}

Many of water's interesting properties stem from its ability to form
complex hydrogen-bonded structures, the description of which requires
an accurate treatment of dispersion interactions. Historically, DFT
has done a poor job of treating these weak interactions. Recent
interest in van der Waals systems has led to the development of a
number of successful approaches aimed at improving DFT's ability to
incorporate dispersion interactions accurately.  These include
post-processing\cite{Misquitta_02, Williams_01} and semi-empirical
methods\cite{Grimme1, Grimme2} as well as hybrid\cite{B3LYP,MPW1B95}
and non-local functionals.\cite{Dion2004,Thonhauser2007} Here, we
focus on the recently developed van der Waals density functional
(vdW-DF), a truly non-local functional which fits seamlessly into DFT.
This functional has been successfully applied to a variety of other
weakly binding systems,\cite{Langreth2009, Cooper2008a, Cooper2008b,
  Thonhauser2006, Hooper2008, Li2009, Bil2011} and is known to hold
promise for improving DFT's description of the hydrogen bonding in
water.\cite{Hamada2010, Hamada2010b, Klimes2010,
  Kelkkanen2009,Wang2011, Li2008} Here, we use it to perform
systematic calculations on small water clusters (H$_2$O)$_n$ with
$n=1-5$ and standard ice $I_h$. Comparison with experiment and
second-order M{\o}ller-Plesset perturbation theory (MP2) calculations
demonstrates that DFT, utilizing vdW-DF, is able to obtain excellent
results for a wide variety of water's properties, even in large,
bulk-like systems.

\section{Computational Details}

\noindent
Our DFT calculations were carried out using the plane-wave
self-consistent field (\textsc{PWscf}) code within the
\textsc{Quantum-Espresso} package,\cite{QE-2009} utilizing ultrasoft
pseudopotentials. For comparison, in addition to vdW-DF we used a
standard local functional (LDA) and a semi-local functional (PBE).
The three functionals all used Slater exchange and Perdew-Wang
correlation\cite{Perdew1992} to describe local exchange and
correlation.  For the PBE functional, the gradient correction approach
of Perdew, Burke, and Ernzerhof was used.\cite{Perdew1996} For vdW-DF,
semi-local exchange was provided by a revised PBE
scheme\cite{Zhang1998} and non-local correlation was provided by
vdW-DF.\cite{Dion2004, Thonhauser2007} Wavefunction and charge-density
cutoffs were 35 Ry and 420 Ry, respectively.  A self-consistency
convergence criterion of at least $1\times 10^{-10}$ Ry was used for
all water cluster calculations and $1\times 10^{-8}$ Ry for ice $I_h$.
Initial structures were relaxed until all force components were less
than $1\times 10^{-5}$~Ry/a.u.\ in the water clusters and $1\times
10^{-4}$~Ry/a.u.\ for ice.  Stress relaxations were carried out in all
ice calculations by relaxing each free lattice parameter until all
stress components were smaller than 1~kbar. To
minimize interactions between periodic images, simulation cells for
the finite systems were sized to ensure a minimum separation of 10
\AA\ between atoms in one cell and atoms in neighboring cells.

MP2 calculations were carried out using the quantum-chemistry package
\textsc{Gaussian},\cite{Gaussian} employing Dunning's augmented,
triple-zeta basis set (aug-cc-pVTZ).  It has been shown that this
basis set is sufficient to yield accurate properties for small water
clusters,\cite{Xantheas1993} and simple testing with the larger
aug-cc-pVQZ basis set confirmed only minimal changes in the properties
of interest here.  Additionally, the MP2 binding energies obtained
here are within 5\% of those reported by Santra et
al.\cite{Santra2007} who extrapolated to the complete basis set. While
counterpoise corrections were deemed negligible for most properties of
interest, we found that they alter the vibrational frequencies by as
much as 6\% in the water dimer.  For this reason, counterpoise
corrections were applied to all MP2 calculations carried out in this
work.  Initial water cluster structures were relaxed until all forces
were less than $4\times 10^{-6}$ Ry/a.u.  Detailed positions for all
optimized structures are provided in the supplementary
materials.

Vibrational frequencies for the MP2 method were calculated directly,
using \textsc{Gaussian}.  For the DFT calculations, forces on all
atoms were calculated when each atom was displaced by $\pm 0.0025$,
$\pm 0.005$, $\pm 0.0075$, and $\pm 0.01$ \AA\ along each of the three
Cartesian directions.  The dynamical matrix was then built by taking
the first derivative of the forces using a nine-point numerical
derivative.  All systems stayed within the harmonic regime for all
displacements employed.

\section{Results}

\subsection{Binding, Structure, and Electric Dipole Moment}

\begin{figure}
  \begin{center} 
    \includegraphics[width=\linewidth]{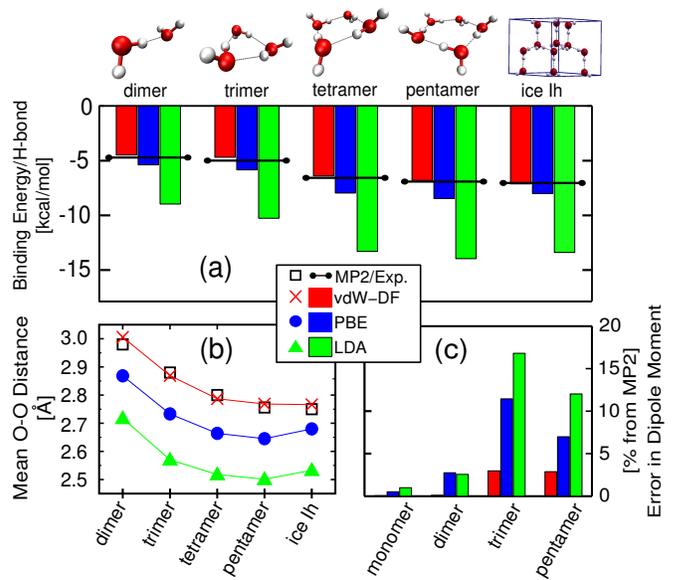}
    \caption{\label{fig:structure}{\bf (a)}~Binding energy per hydrogen
      bond for each of the multimers and ice $I_h$.  The horizontal
      black lines correspond to MP2 calculations for the clusters and
      experiment for ice.\cite{Petrenko1999} {\bf (b)}~Average
      oxygen-oxygen distance in each system.  The black boxes represent
      experimental values.\cite{Liu1996, Bergmann2007} {\bf (c)}~Percent
      error relative to MP2 calculations of the dipole moment. The
      dipole moment of the tetramer is not shown since it is zero by the
      S$_4$ symmetry of the complex.}
  \end{center}
\end{figure}

\noindent
We start by presenting results for the binding energies of the water
clusters (H$_2$O)$_n$ with $n=2-5$, and standard, hexagonal ice $I_h$.
The results are shown in Fig.~\ref{fig:structure}(a), along with MP2
values for the water clusters and the experimental value for ice
$I_h$.\cite{Petrenko1999} Also shown for reference are values from two
of the most popular local (LDA) and semi-local (PBE) functionals in
use today.  As can be seen in the figure, the LDA and PBE functionals
exhibit a distinct overbinding.  This is also evident in
Fig.~\ref{fig:structure}(b), which shows the average oxygen-oxygen
distance compared with experiment for both the clusters\cite{Liu1996}
and ice $I_h$.\cite{Bergmann2007} The vdW-DF values are in excellent
agreement with experiment and show a clear improvement over LDA and
PBE, which draw the oxygen atoms too close together, consistent with
the overbinding seen in Fig.~\ref{fig:structure}(a).

Interestingly, a recent study by Wang et al.\cite{Wang2011} has
pointed out that, despite showing a substantial improvement over
other functionals when calculating the self-diffusion and density
of liquid water, vdW-DF tends to
understructure the liquid.  This understructuring manifests itself as
a lowered density in the second coordination shell of the
oxygen-oxygen radial distribution function.  Wang et al. point out
that this understructuring is likely an artifact of the chosen
semi-local piece of the exchange functional, rather than a problem
within the non-local correlation functional of vdW-DF.  Many groups
are investigating the effects of different exchange functionals on
vdW-DF and it will be interesting to see if the understructuring
problem can be solved.

Another important quantity is the bulk modulus---a property that depends
on the curvature of the energy surface when the system is subjected to
isotropic expansion or contraction.  In a recent study,\cite{Hamada2010}
vdW-DF was used to calculate the bulk modulus of ice $I_h$.  This study
found vdW-DF to be in good agreement with the experimental value and
serves as an independent extension of the findings presented here.

To probe the accuracy of the electronic structure itself, the dipole
moments were calculated for the finite systems.
Figure~\ref{fig:structure}(c) shows the errors obtained (relative to
MP2 calculations) for the three functionals studied here.
Experimental results could not be found for clusters larger than
$n=2$,\cite{Gregory1998} but MP2 results agree with available
experimental results to within 1\%, so MP2 is used as the reference
throughout Fig.~\ref{fig:structure}(c). Detailed values for all dipole
moments are presented in the supplementary materials.  Again, results
obtained from vdW-DF are in excellent agreement with these high-level
quantum-chemistry calculations, while LDA and PBE show substantial
errors.

\begin{figure}
  \begin{center}
    \includegraphics[width=\linewidth]{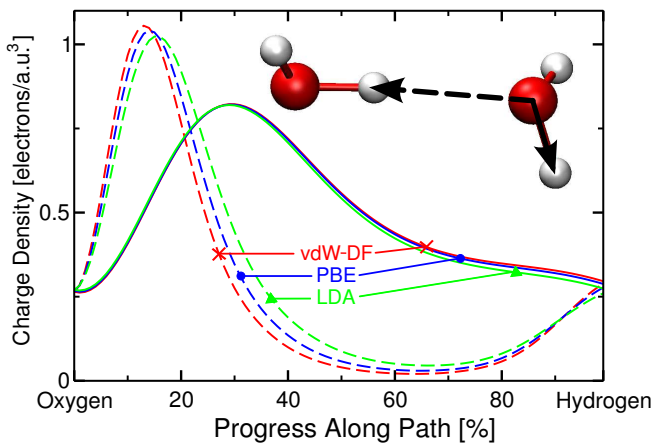}
    \caption{\label{fig:density}Calculated valence charge density for
      the water dimer along the lines between the indicated oxygen and
      hydrogens through both the covalent bond (solid lines) and the
      hydrogen bond (dashed lines) for the three different functionals.}
  \end{center}
\end{figure}

One can understand the better performance of vdW-DF since it is a
non-local functional, but the physical ramifications of this
distinction can be seen in Fig.~\ref{fig:density}. The figure shows
the calculated charge density along a line from oxygen to hydrogen
through both the covalent bond (solid lines) and the hydrogen bond
(dashed lines). Not surprising, the charge densities corresponding to
the covalent bond for each functional are nearly
indistinguishable. However, when looking at the curves corresponding
to the charge density in the hydrogen bond, the cause for the
overbinding of LDA and PBE---evident throughout
Fig.~\ref{fig:structure}---is revealed.  Both PBE, and to an even
greater extent LDA, shift the peak of charge density from the oxygen
toward the hydrogen, resulting in an increase in the covalent
character of the bond.  (Note that, in the critical region between
20\% and 60\% along the path, LDA overestimates the charge density by
as much as a factor of 2.) This, in turn, strengthens the bond and
creates an overbinding that permeates all results calculated with
these functionals.

\subsection{Vibrational Frequencies}

\begin{figure}
  \begin{center}
    \includegraphics[width=\linewidth]{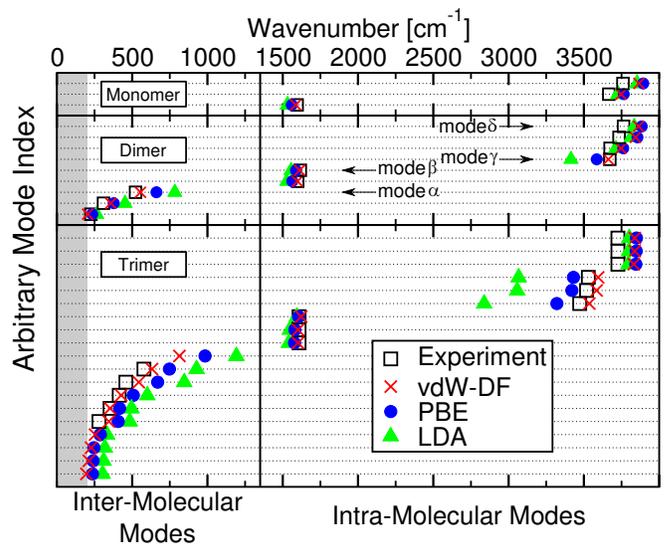}
    \caption{\label{fig:frequencies}Vibrational frequencies calculated
      with the LDA, PBE, and vdW-DF functionals compared to
      experimental values for the water monomer,\cite{Ceponkus2008}
      dimer,\cite{Ceponkus2008} and trimer.\cite{Tremblay2010} The
      vertical line denotes a separation between inter-molecular
      dominated and intra-molecular dominated modes.  The region
      between 0 cm$^{-1}$ and 200 cm$^{-1}$ (shaded area) was not
      included in the analysis.  A detailed analysis of the dimer
      modes labeled mode $\alpha$, $\beta$, $\gamma$, and $\delta$ is
      given in Figs.~\ref{fig:breakdown_ac} and
      \ref{fig:breakdown_bd}.  A full tabulation of mode frequencies
      for all methods can be found in the supplementary material.}
  \end{center}
\end{figure}

\noindent
Sit and Marzari\cite{Sit2005} calculated the vibrational frequencies
of the water dimer using the PBE functional. They found that PBE
performed reasonably well, but significant discrepancies from
experiment were found for some modes, particularly low-frequency,
inter-molecular modes. They postulated, as have
others,\cite{Kelkkanen2009} that the problem arose at least in part
from the inability of local and semi-local functionals to correctly
treat the hydrogen bond. To extend this investigation, we performed
similar frequency calculations for all the water clusters---monomer
through pentamer---using all three functionals and MP2.  (Our results
for the dimer using PBE are very similar to those of Sit and Marzari.)
The results for the larger water clusters exhibit trends similar to
those found in the dimer.  Figure~\ref{fig:frequencies} shows our
DFT-calculated and available experimentally-determined vibrational
frequencies for the monomer,\cite{Ceponkus2008}
dimer,\cite{Ceponkus2008} and trimer;\cite{Tremblay2010} reliable
experimental results for larger clusters could not be found. Results
for all clusters compared with MP2 and the frequencies of ice $I_h$
calculated with the three functionals are collected in the
supplementary materials. It should be noted that the frequencies
calculated in this work were obtained using the harmonic approximation
and no corrections were made for anharmonic effects. Such effects are
known to be present in experimental results and can be influential in
setting the precise frequency of various oscillations. Calculations
including such effects have recently been reported at the PBE and PBE0
level,\cite{Zhang2011} but a direct comparison is not possible since
the frequencies reported are for the deuterated monomer and dimer.
Typically, the changes induced by anharmonic effects are on the order
of a few percent and fair comparisons can still be made to
experiment. Nevertheless, in this work, truly quantitative comparisons
of DFT results were done with respect to the harmonic approximation
within MP2.

As can be seen in Fig.~\ref{fig:frequencies}, all DFT methods get
similar results for the monomer frequencies---not surprising since
this is a system consisting solely of covalent bonds.  Significant
deviations occur in some modes for the larger clusters, however.
These deviations are worse in the low-frequency region where
inter-molecular interactions dominate. Most of the
intra-molecular-dominated frequencies are obtained satisfactorily with
all three functionals, but a few show significant error in LDA and
PBE.  Even within the intra-molecular modes there are oscillations
which substantially change the hydrogen-bond geometry, so a proper
treatment of non-local interactions is necessary to obtain
quantitatively correct frequencies. Overall, vdW-DF shows a consistent
improvement for a majority of the modes.  A full tabulation of modes
for all methods and systems (including ice $I_h$) is given in the
supplementary materials.

\begin{figure}
\includegraphics[width=0.85\linewidth]{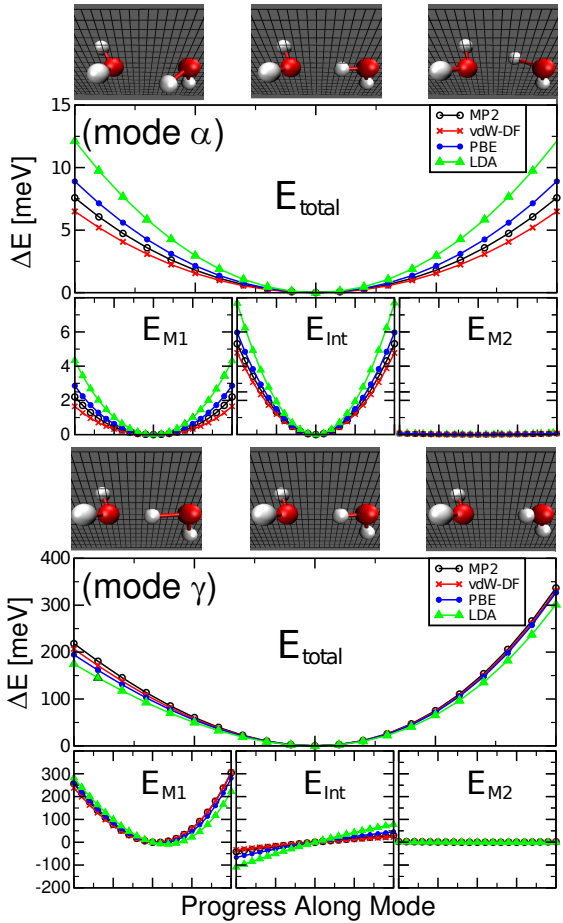}
\caption{\label{fig:breakdown_ac} Energy changes according to
  Eq.~(\ref{deltaE}) as the water dimer is forced along the normal
  modes marked $\alpha$ and $\gamma$ in
  Fig.~\ref{fig:frequencies}. The x-axes are on the same scale and
  represent a dimensionless progress variable along the mode. The
  insets above show the motion of the water molecules for each
  particular mode.  The grids lie in a plane containing the hydrogen
  bonding O--H--O in the equilibrium position.  In this figure, the
  hydrogen donor molecule is designated M1 and the hydrogen
  acceptor molecule is designated M2.}
\end{figure}

\begin{figure}
\includegraphics[width=0.85\linewidth]{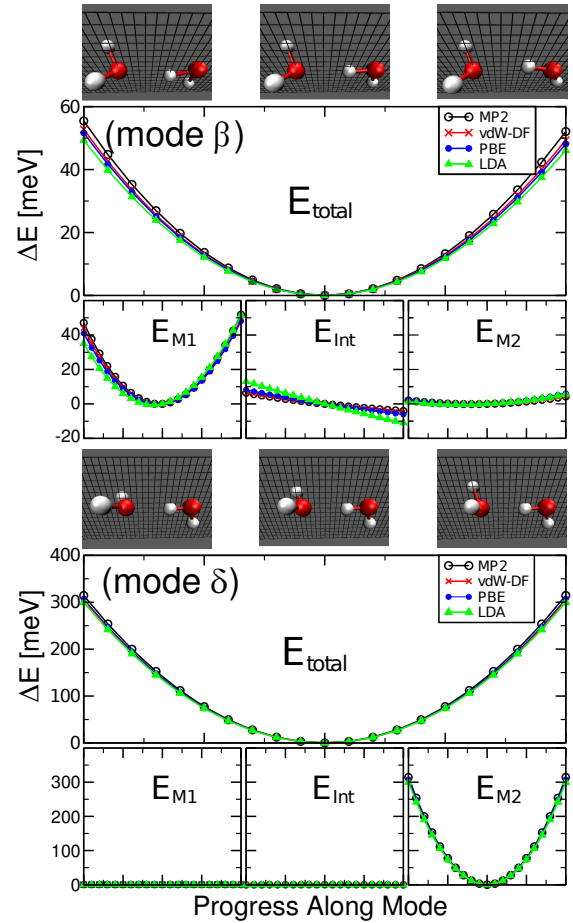}
\caption{\label{fig:breakdown_bd} Same as Fig.~\ref{fig:breakdown_ac},
  except here shown for the modes marked $\beta$ and $\delta$
  in Fig.~\ref{fig:frequencies}.}
\end{figure}

While some modes are well reproduced with all functionals, others show
a distinct spread in frequencies for different functionals. This is to
be expected for inter-molecular modes, but it is somewhat surprising
for the intra-molecular modes. This spread is caused largely by the
different functionals' varying ability to accurately represent the
hydrogen bond, as can be seen by a deeper analysis of some
representative modes of the dimer system.  Four such modes---labeled
mode $\alpha$, mode $\beta$, mode $\gamma$, and mode
$\delta$---are marked in Fig.~\ref{fig:frequencies}.  Modes
$\alpha$ and $\gamma$ show a large spread in frequency for
different functionals, while modes $\beta$ and $\delta$ exhibit
much less variation.  We analyze these modes in greater detail as
follows.

The change in energy of any dimer configuration relative to the
equilibrium configuration can be written as
\begin{equation}
\Delta E_{\text{total}} = \Delta E_{\text{M1}} +
\Delta E_{\text{M2}} + \Delta E_{\text{int}}\;,
\label{deltaE}
\end{equation}
where $\Delta E_{\text{M1}}$ and $\Delta E_{\text{M2}}$ are
the strain energies within monomers 1 and 2, respectively, and $\Delta
E_{\text{int}}$ is the difference in their interaction energy. As the
atoms are moved along a certain mode, the individual contributions in
Eq.~(\ref{deltaE}) can increase, decrease, or remain unchanged,
depending on whether they enhance, inhibit, or are irrelevant to the
change in energy for small excitations of that particular mode.  The
change in energy coming from these contributions relative to the
overall energy change determines how much each energy term contributes
to the vibrational frequency.

Figures~\ref{fig:breakdown_ac} and \ref{fig:breakdown_bd} show the
contributions of the various energy terms in Eq.~(\ref{deltaE}) as the
dimer is displaced in both directions along the normal modes
$\alpha$, $\beta$, $\gamma$, and $\delta$.  This was done in
step sizes of 0.01 times the (normalized) eigenvectors, to a maximum
value of 0.1 times the eigenvectors.  Mode $\alpha$ is comprised
mainly of an out-of-plane oscillatory motion of the hydrogen donor
water molecule.  Oscillations along this mode clearly change the
geometry of the hydrogen bond.  Since the overall energy change upon
movement along this mode is relatively small, the interaction energy
makes up a significant portion of the total restoring force. As the
different functionals have a varying degree of success in describing
the interaction energy, a significant spread in $E_{\text{total}}$ and
the frequencies results for different functionals. The motion for mode
$\gamma$ consists mainly of a symmetric stretch of the hydrogen
donor molecule.  Again, this substantially changes the interaction
energy as the dimer oscillates along the mode, resulting in a spread
in $E_{\text{total}}$ and the frequencies. On the other hand, mode
$\beta$ in Fig.~\ref{fig:breakdown_bd} is comprised mostly of an
angle flex in the hydrogen donor molecule.  As with mode $\alpha$,
oscillations along this mode change the hydrogen bond angle and, thus,
the interaction energy. Unlike mode $\alpha$, however, the
intra-molecular motions in this mode cause the total energy to change
by a relatively large amount.  In this case it is enough to swamp the
effect from the relatively weak energy change stemming from the
changing hydrogen bond geometry. As a result, all functionals find
similar frequencies for this mode. Mode $\delta$ corresponds to an
asymmetric stretch of the hydrogen acceptor molecule.  This has
virtually no effect on the hydrogen-bond geometry.  The frequency of
this mode is governed almost entirely by the intra-molecular strain
energy in the hydrogen acceptor molecule, with the interaction energy
playing virtually no role. Thus, all functionals predict very similar
frequencies for this mode. Qualitatively, for the four modes discussed
the spread in $E_{\text{total}}$ is proportional to the relative
spread in frequency in Fig.~\ref{fig:frequencies}.

\begin{figure}
  \begin{center}
    \includegraphics[width=\linewidth]{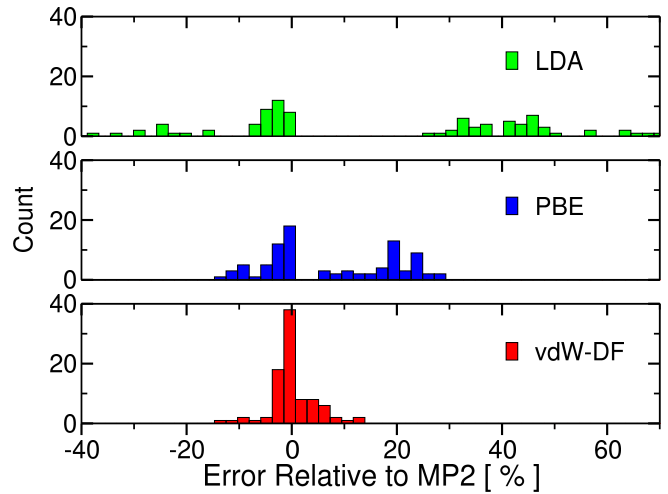}
    \caption{\label{fig:histogram}Histograms of the percent error
    relative to MP2 calculations over all the frequencies of the
    clusters. The horizontal axis gives the percent error and the
    vertical axis gives the number of vibrational frequencies that had
    that error.}    
  \end{center}
\end{figure}

To conclude our study of the vibrational properties, the vdW-DF
frequencies of all clusters, including the tetramer and the pentamer,
were quantitatively compared to MP2 calculations.  The results of this
analysis are shown in Fig.~\ref{fig:histogram} in the form of
histograms of the errors (in percent relative to MP2), analyzing 90
modes in total (as in Fig.~\ref{fig:frequencies}, modes with a
frequency less than 200 cm$^{-1}$ were not included). Again, a
complete listing of frequencies for these modes can be found in the
supplementary material. In the figure a negative percent error simply
means the calculated value was \emph{below} the MP2 value.  LDA shows
a distinct bimodal distribution in errors.  These correspond to a
systematic overestimation of the frequency of low-frequency,
inter-molecular modes and less pronounced underestimation of the
high-frequency, intra-molecular modes relative to MP2.  PBE has a much
tighter error range but still shows the bimodal distribution evident
in the LDA results, and for the same reason.  As discussed above, both
functionals form an artificially strong hydrogen bond. This---together
with the fact that inter-molecular interactions tend to enhance the
overall restoring force for inter-molecular modes and to weaken it in
intra-molecular modes---results in the bimodal distribution in the LDA
and PBE results.  The vdW-DF plot shows an error spread that is much
less severe and does not exhibit the bimodal distribution
characteristic of systematic hydrogen-bond overestimation.  From
Figs.~\ref{fig:frequencies} and \ref{fig:histogram} it is clear that
vdW-DF systematically improves upon commonly used functionals and
demonstrates great promise to accurately reproduce vibrational
frequencies of water.

\section{Conclusions}

\noindent
We have used density functional theory incorporating the van der Waals
density functional to calculate a number of properties of small water
clusters (H$_2$O)$_n$ with $n=1-5$ and hexagonal ice $I_h$. Our
results show excellent agreement, both with experiment and high-level
quantum-chemistry calculations.  Since DFT is capable of calculating
systems with a thousand atoms or more, this implies that accurate
quantum-mechanics can now be applied to bulk-like water.  This
capability should rapidly advance our detailed knowledge and
understanding of bulk water behavior.

Toward this aim, the authors, working in collaboration with the
developers of the free, open-source \textsc{Quantum-Espresso}
package,\cite{QE-2009} have implemented an
efficient\cite{Roman-Perez2009} vdW-DF functional in the newest official
release.  We strongly encourage the testing and application of this, and
the newer vdW-DF2\cite{Lee2010} functional (also implemented in
\textsc{Quantum-Espresso}) in water-based systems and, indeed, in any
system where hydrogen bonding is expected to play a critical role.

\section*{Acknowledgements}

\noindent
B.\,K. would like to thank Drs.\ N. Holzwarth and B. Kerr for many
helpful discussions. The authors would also like to thank G. Galli for
providing the initial ice $I_h$ structure.  Partial
funding from Oak Ridge Associated Universities (ORAU) is acknowledged.

%


\begin{thebibliography}{10}%
\makeatletter
\providecommand \@ifxundefined [1]{%
 \ifx #1\undefined \expandafter \@firstoftwo
 \else \expandafter \@secondoftwo
\fi
}%
\providecommand \@ifnum [1]{%
 \ifnum #1\expandafter \@firstoftwo
 \else \expandafter \@secondoftwo
\fi
}%
\providecommand \enquote [1]{``#1''}%
\providecommand \bibnamefont  [1]{#1}%
\providecommand \bibfnamefont [1]{#1}%
\providecommand \citenamefont [1]{#1}%
\providecommand\href[0]{\@sanitize\@href}%
\providecommand\@href[1]{\endgroup\@@startlink{#1}\endgroup\@@href}%
\providecommand\@@href[1]{#1\@@endlink}%
\providecommand \@sanitize [0]{\begingroup\catcode`\&12\catcode`\#12\relax}%
\@ifxundefined \pdfoutput {\@firstoftwo}{%
 \@ifnum{\z@=\pdfoutput}{\@firstoftwo}{\@secondoftwo}%
}{%
 \providecommand\@@startlink[1]{\leavevmode}%
 \providecommand\@@endlink[0]{}%
}{%
 \providecommand\@@startlink[1]{%
  \leavevmode
  \pdfstartlink
   attr{/Border[0 0 1 ]/H/I/C[0 1 1]}%
   user{/Subtype/Link/A<</Type/Action/S/URI/URI(#1)>>}%
  \relax
 }%
 \providecommand\@@endlink[0]{\pdfendlink}%
}%
\providecommand \url  [0]{\begingroup\@sanitize \@url }%
\providecommand \@url [1]{\endgroup\@href {#1}{\urlprefix}}%
\providecommand \urlprefix [0]{URL }%
\providecommand \Eprint[0]{\href }%
\@ifxundefined \urlstyle {%
  \providecommand \doi [1]{doi:\discretionary{}{}{}#1}%
}{%
  \providecommand \doi [0]{doi:\discretionary{}{}{}\begingroup
  \urlstyle{rm}\Url }%
}%
\providecommand \doibase [0]{http://dx.doi.org/}%
\providecommand \Doi[1]{\href{\doibase#1}}%
\providecommand \bibAnnote [3]{%
  \BibitemShut{#1}%
  \begin{quotation}\noindent
    \textsc{Key:}\ #2\\\textsc{Annotation:}\ #3%
  \end{quotation}%
}%
\providecommand \bibAnnoteFile [2]{%
  \IfFileExists{#2}{\bibAnnote {#1} {#2} {\input{#2}}}{}%
}%
\providecommand \typeout [0]{\immediate \write \m@ne }%
\providecommand \selectlanguage [0]{\@gobble}%
\providecommand \bibinfo [0]{\@secondoftwo}%
\providecommand \bibfield [0]{\@secondoftwo}%
\providecommand \translation [1]{[#1]}%
\providecommand \BibitemOpen[0]{}%
\providecommand \bibitemStop [0]{}%
\providecommand \bibitemNoStop [0]{.\EOS\space}%
\providecommand \EOS [0]{\spacefactor3000\relax}%
\providecommand \BibitemShut [1]{\csname bibitem#1\endcsname}%
\bibitem{Guillot2002}%
  \BibitemOpen
  \bibfield{author}{%
  \bibinfo {author} {\bibfnamefont{B.}~\bibnamefont{Guillot}},\ }%
  \bibfield{journal}{%
  \bibinfo {journal} {J. Mol. Liq.}\ }%
  \textbf{\bibinfo {volume} {101}},\ \bibinfo {pages} {219} (\bibinfo {year}
  {2002})%
  \bibAnnoteFile{NoStop}{Guillot2002}%
\bibitem{Abascal2009}%
  \BibitemOpen
  \bibfield{author}{%
  \bibinfo {author} {\bibfnamefont{J.~L.~F.}\ \bibnamefont{Abascal}}, \bibinfo
  {author} {\bibfnamefont{E.}~\bibnamefont{Sanz}},\ and\ \bibinfo {author}
  {\bibfnamefont{C.}~\bibnamefont{Vega}},\ }%
  \bibfield{journal}{%
  \bibinfo {journal} {Phys. Chem. Chem. Phys.}\ }%
  \textbf{\bibinfo {volume} {11}},\ \bibinfo {pages} {556} (\bibinfo {year}
  {2009})%
  \bibAnnoteFile{NoStop}{Abascal2009}%
\bibitem{Sanz2004}%
  \BibitemOpen
  \bibfield{author}{%
  \bibinfo {author} {\bibfnamefont{E.}~\bibnamefont{Sanz}}, \bibinfo {author}
  {\bibfnamefont{C.}~\bibnamefont{Vega}}, \bibinfo {author}
  {\bibfnamefont{J.~L.~F.}\ \bibnamefont{Abascal}},\ and\ \bibinfo {author}
  {\bibfnamefont{L.~G.}\ \bibnamefont{MacDowell}},\ }%
  \bibfield{journal}{%
  \bibinfo {journal} {J. Chem. Phys.}\ }%
  \textbf{\bibinfo {volume} {121}},\ \bibinfo {pages} {1165} (\bibinfo {year}
  {2004})%
  \bibAnnoteFile{NoStop}{Sanz2004}%
\bibitem{Xantheas1993}%
  \BibitemOpen
  \bibfield{author}{%
  \bibinfo {author} {\bibfnamefont{S.~S.}\ \bibnamefont{Xantheas}}\ and\
  \bibinfo {author} {\bibfnamefont{T.~H.}\ \bibnamefont{Dunning},
  \bibfnamefont{Jr.}},\ }%
  \bibfield{journal}{%
  \bibinfo {journal} {J. Chem. Phys.}\ }%
  \textbf{\bibinfo {volume} {99}},\ \bibinfo {pages} {8774} (\bibinfo {year}
  {1993})%
  \bibAnnoteFile{NoStop}{Xantheas1993}%
\bibitem{Xantheas2002}%
  \BibitemOpen
  \bibfield{author}{%
  \bibinfo {author} {\bibfnamefont{S.~S.}\ \bibnamefont{Xantheas}}, \bibinfo
  {author} {\bibfnamefont{C.~J.}\ \bibnamefont{Burnham}},\ and\ \bibinfo
  {author} {\bibfnamefont{R.~J.}\ \bibnamefont{Harrison}},\ }%
  \bibfield{journal}{%
  \bibinfo {journal} {J. Chem. Phys.}\ }%
  \textbf{\bibinfo {volume} {116}},\ \bibinfo {pages} {1493} (\bibinfo {year}
  {2002})%
  \bibAnnoteFile{NoStop}{Xantheas2002}%
\bibitem{Gregory1997}%
  \BibitemOpen
  \bibfield{author}{%
  \bibinfo {author} {\bibfnamefont{J.~K.}\ \bibnamefont{Gregory}}, \bibinfo
  {author} {\bibfnamefont{D.~C.}\ \bibnamefont{Clary}}, \bibinfo {author}
  {\bibfnamefont{K.}~\bibnamefont{Liu}}, \bibinfo {author}
  {\bibfnamefont{M.~G.}\ \bibnamefont{Brown}},\ and\ \bibinfo {author}
  {\bibfnamefont{R.~J.}\ \bibnamefont{Saykally}},\ }%
  \bibfield{journal}{%
  \bibinfo {journal} {Science}\ }%
  \textbf{\bibinfo {volume} {275}},\ \bibinfo {pages} {814} (\bibinfo {year}
  {1997})%
  \bibAnnoteFile{NoStop}{Gregory1997}%
\bibitem{Shields2010}%
  \BibitemOpen
  \bibfield{author}{%
  \bibinfo {author} {\bibfnamefont{R.~M.}\ \bibnamefont{Shields}}, \bibinfo
  {author} {\bibfnamefont{B.}~\bibnamefont{Temelso}}, \bibinfo {author}
  {\bibfnamefont{K.~A.}\ \bibnamefont{Archer}}, \bibinfo {author}
  {\bibfnamefont{T.~E.}\ \bibnamefont{Morrell}},\ and\ \bibinfo {author}
  {\bibfnamefont{G.~C.}\ \bibnamefont{Shields}},\ }%
  \bibfield{journal}{%
  \bibinfo {journal} {J. Phys. Chem. A}\ }%
  \textbf{\bibinfo {volume} {114}},\ \bibinfo {pages} {11725} (\bibinfo {year}
  {2010})%
  \bibAnnoteFile{NoStop}{Shields2010}%
\bibitem{Head-Gordon2008}%
  \BibitemOpen
  \bibfield{author}{%
  \bibinfo {author} {\bibfnamefont{M.}~\bibnamefont{Head-Gordon}}\ and\
  \bibinfo {author} {\bibfnamefont{E.}~\bibnamefont{Artacho}},\ }%
  \bibfield{journal}{%
  \bibinfo {journal} {Physics Today}\ }%
  \textbf{\bibinfo {volume} {61}},\ \bibinfo {pages} {58} (\bibinfo {year}
  {2008})%
  \bibAnnoteFile{NoStop}{Head-Gordon2008}%
\bibitem{schwegler:5400}%
  \BibitemOpen
  \bibfield{author}{%
  \bibinfo {author} {\bibfnamefont{E.}~\bibnamefont{Schwegler}}, \bibinfo
  {author} {\bibfnamefont{J.~C.}\ \bibnamefont{Grossman}}, \bibinfo {author}
  {\bibfnamefont{F.}~\bibnamefont{Gygi}},\ and\ \bibinfo {author}
  {\bibfnamefont{G.}~\bibnamefont{Galli}},\ }%
  \bibfield{journal}{%
  \Doi{10.1063/1.1782074}{\bibinfo {journal} {J. Chem. Phys.}}\ }%
  \textbf{\bibinfo {volume} {121}},\ \bibinfo {pages} {5400} (\bibinfo {year}
  {2004})%
  \bibAnnoteFile{NoStop}{schwegler:5400}%
\bibitem{fernandez-serra:11136}%
  \BibitemOpen
  \bibfield{author}{%
  \bibinfo {author} {\bibfnamefont{M.~V.}\ \bibnamefont{Fern\'{a}1ndez-Serra}}\
  and\ \bibinfo {author} {\bibfnamefont{E.}~\bibnamefont{Artacho}},\ }%
  \bibfield{journal}{%
  \Doi{10.1063/1.1813431}{\bibinfo {journal} {J. Chem. Phys.}}\ }%
  \textbf{\bibinfo {volume} {121}},\ \bibinfo {pages} {11136} (\bibinfo {year}
  {2004})%
  \bibAnnoteFile{NoStop}{fernandez-serra:11136}%
\bibitem{lee:154507}%
  \BibitemOpen
  \bibfield{author}{%
  \bibinfo {author} {\bibfnamefont{H.-S.}\ \bibnamefont{Lee}}\ and\ \bibinfo
  {author} {\bibfnamefont{M.~E.}\ \bibnamefont{Tuckerman}},\ }%
  \bibfield{journal}{%
  \Doi{10.1063/1.2354158}{\bibinfo {journal} {J. Chem. Phys.}}\ }%
  \textbf{\bibinfo {volume} {125}},\ \bibinfo {eid} {154507} (\bibinfo {year}
  {2006})%
  \bibAnnoteFile{NoStop}{lee:154507}%
\bibitem{guidon:214104}%
  \BibitemOpen
  \bibfield{author}{%
  \bibinfo {author} {\bibfnamefont{M.}~\bibnamefont{Guidon}}, \bibinfo {author}
  {\bibfnamefont{F.}~\bibnamefont{Schiffmann}}, \bibinfo {author}
  {\bibfnamefont{J.}~\bibnamefont{Hutter}},\ and\ \bibinfo {author}
  {\bibfnamefont{J.}~\bibnamefont{VandeVondele}},\ }%
  \bibfield{journal}{%
  \Doi{10.1063/1.2931945}{\bibinfo {journal} {J. Chem. Phys.}}\ }%
  \textbf{\bibinfo {volume} {128}},\ \bibinfo {eid} {214104} (\bibinfo {year}
  {2008})%
  \bibAnnoteFile{NoStop}{guidon:214104}%
\bibitem{Sit2005}%
  \BibitemOpen
  \bibfield{author}{%
  \bibinfo {author} {\bibfnamefont{P.~H.-L.}\ \bibnamefont{Sit}}\ and\ \bibinfo
  {author} {\bibfnamefont{N.}~\bibnamefont{Marzari}},\ }%
  \bibfield{journal}{%
  \bibinfo {journal} {J. Chem. Phys.}\ }%
  \textbf{\bibinfo {volume} {122}},\ \bibinfo {pages} {204510} (\bibinfo {year}
  {2005})%
  \bibAnnoteFile{NoStop}{Sit2005}%
\bibitem{Misquitta_02}%
  \BibitemOpen
  \bibfield{author}{%
  \bibinfo {author} {\bibfnamefont{A.~J.}\ \bibnamefont{Misquitta}}\ and\
  \bibinfo {author} {\bibfnamefont{K.}~\bibnamefont{Szalewicz}},\ }%
  \bibfield{journal}{%
  \bibinfo {journal} {Chem. Phys. Lett.}\ }%
  \textbf{\bibinfo {volume} {357}},\ \bibinfo {pages} {301} (\bibinfo {year}
  {2002})%
  \bibAnnoteFile{NoStop}{Misquitta_02}%
\bibitem{Williams_01}%
  \BibitemOpen
  \bibfield{author}{%
  \bibinfo {author} {\bibfnamefont{H.~L.}\ \bibnamefont{Williams}}\ and\
  \bibinfo {author} {\bibfnamefont{C.~F.}\ \bibnamefont{Chabalowski}},\ }%
  \bibfield{journal}{%
  \bibinfo {journal} {J. Phys. Chem. A}\ }%
  \textbf{\bibinfo {volume} {105}},\ \bibinfo {pages} {646} (\bibinfo {year}
  {2001})%
  \bibAnnoteFile{NoStop}{Williams_01}%
\bibitem{Grimme1}%
  \BibitemOpen
  \bibfield{author}{%
  \bibinfo {author} {\bibfnamefont{S.}~\bibnamefont{Grimme}},\ }%
  \bibfield{journal}{%
  \bibinfo {journal} {J. Comp. Chem.}\ }%
  \textbf{\bibinfo {volume} {25}},\ \bibinfo {pages} {1463} (\bibinfo {year}
  {2004})%
  \bibAnnoteFile{NoStop}{Grimme1}%
\bibitem{Grimme2}%
  \BibitemOpen
  \bibfield{author}{%
  \bibinfo {author} {\bibfnamefont{S.}~\bibnamefont{Grimme}},\ }%
  \bibfield{journal}{%
  \bibinfo {journal} {J. Comp. Chem.}\ }%
  \textbf{\bibinfo {volume} {27}},\ \bibinfo {pages} {1787} (\bibinfo {year}
  {2006})%
  \bibAnnoteFile{NoStop}{Grimme2}%
\bibitem{B3LYP}%
  \BibitemOpen
  \bibfield{author}{%
  \bibinfo {author} {\bibfnamefont{P.~J.}\ \bibnamefont{Stephens}}, \bibinfo
  {author} {\bibfnamefont{F.~J.}\ \bibnamefont{Devlin}}, \bibinfo {author}
  {\bibfnamefont{C.~F.}\ \bibnamefont{Chabalowski}},\ and\ \bibinfo {author}
  {\bibfnamefont{M.~J.}\ \bibnamefont{Frisch}},\ }%
  \bibfield{journal}{%
  \bibinfo {journal} {J. Phys. Chem.}\ }%
  \textbf{\bibinfo {volume} {98}},\ \bibinfo {pages} {11623} (\bibinfo {year}
  {1994})%
  \bibAnnoteFile{NoStop}{B3LYP}%
\bibitem{MPW1B95}%
  \BibitemOpen
  \bibfield{author}{%
  \bibinfo {author} {\bibfnamefont{Y.}~\bibnamefont{Zhao}}\ and\ \bibinfo
  {author} {\bibfnamefont{D.~G.}\ \bibnamefont{Truhlar}},\ }%
  \bibfield{journal}{%
  \bibinfo {journal} {J. Phys. Chem. A}\ }%
  \textbf{\bibinfo {volume} {108}},\ \bibinfo {pages} {6908} (\bibinfo {year}
  {2004})%
  \bibAnnoteFile{NoStop}{MPW1B95}%
\bibitem{Dion2004}%
  \BibitemOpen
  \bibfield{author}{%
  \bibinfo {author} {\bibfnamefont{M.}~\bibnamefont{Dion}}, \bibinfo {author}
  {\bibfnamefont{H.}~\bibnamefont{Rydberg}}, \bibinfo {author}
  {\bibfnamefont{E.}~\bibnamefont{Schr\"{o}der}}, \bibinfo {author}
  {\bibfnamefont{D.~C.}\ \bibnamefont{Langreth}},\ and\ \bibinfo {author}
  {\bibfnamefont{B.}~\bibnamefont{Lundqvist}},\ }%
  \bibfield{journal}{%
  \bibinfo {journal} {Phys. Rev. Lett.}\ }%
  \textbf{\bibinfo {volume} {92}},\ \bibinfo {pages} {246401} (\bibinfo {year}
  {2004})%
  \bibAnnoteFile{NoStop}{Dion2004}%
\bibitem{Thonhauser2007}%
  \BibitemOpen
  \bibfield{author}{%
  \bibinfo {author} {\bibfnamefont{T.}~\bibnamefont{Thonhauser}}, \bibinfo
  {author} {\bibfnamefont{V.~R.}\ \bibnamefont{Cooper}}, \bibinfo {author}
  {\bibfnamefont{S.}~\bibnamefont{Li}}, \bibinfo {author}
  {\bibfnamefont{A.}~\bibnamefont{Puzder}}, \bibinfo {author}
  {\bibfnamefont{P.}~\bibnamefont{Hyldgaard}},\ and\ \bibinfo {author}
  {\bibfnamefont{D.~C.}\ \bibnamefont{Langreth}},\ }%
  \bibfield{journal}{%
  \bibinfo {journal} {Phys. Rev. B}\ }%
  \textbf{\bibinfo {volume} {76}},\ \bibinfo {pages} {125112} (\bibinfo {year}
  {2007})%
  \bibAnnoteFile{NoStop}{Thonhauser2007}%
\bibitem{Langreth2009}%
  \BibitemOpen
  \bibfield{author}{%
  \bibinfo {author} {\bibfnamefont{D.~C.}\ \bibnamefont{Langreth}}, \bibinfo
  {author} {\bibfnamefont{B.~I.}\ \bibnamefont{Lundqvist}}, \bibinfo {author}
  {\bibfnamefont{S.~D.}\ \bibnamefont{Chakarova-K{\"a}ck}}, \bibinfo {author}
  {\bibfnamefont{V.~R.}\ \bibnamefont{Cooper}}, \bibinfo {author}
  {\bibfnamefont{M.}~\bibnamefont{Dion}}, \bibinfo {author}
  {\bibfnamefont{P.}~\bibnamefont{Hyldgaard}}, \bibinfo {author}
  {\bibfnamefont{A.}~\bibnamefont{Kelkkanen}}, \bibinfo {author}
  {\bibfnamefont{J.}~\bibnamefont{Kleis}}, \bibinfo {author}
  {\bibfnamefont{L.}~\bibnamefont{Kong}}, \bibinfo {author}
  {\bibfnamefont{S.}~\bibnamefont{Li}}, \bibinfo {author}
  {\bibfnamefont{P.~G.}\ \bibnamefont{Moses}}, \bibinfo {author}
  {\bibfnamefont{E.}~\bibnamefont{Murray}}, \bibinfo {author}
  {\bibfnamefont{A.}~\bibnamefont{Puzder}}, \bibinfo {author}
  {\bibfnamefont{H.}~\bibnamefont{Rydberg}}, \bibinfo {author}
  {\bibfnamefont{E.}~\bibnamefont{Schr{\"o}der}},\ and\ \bibinfo {author}
  {\bibfnamefont{T.}~\bibnamefont{Thonhauser}},\ }%
  \bibfield{journal}{%
  \bibinfo {journal} {J. Phys.: Condensed Matter}\ }%
  \textbf{\bibinfo {volume} {21}},\ \bibinfo {pages} {084203} (\bibinfo {year}
  {2009})%
  \bibAnnoteFile{NoStop}{Langreth2009}%
\bibitem{Cooper2008a}%
  \BibitemOpen
  \bibfield{author}{%
  \bibinfo {author} {\bibfnamefont{V.~R.}\ \bibnamefont{Cooper}}, \bibinfo
  {author} {\bibfnamefont{T.}~\bibnamefont{Thonhauser}}, \bibinfo {author}
  {\bibfnamefont{A.}~\bibnamefont{Puzder}}, \bibinfo {author}
  {\bibfnamefont{E.}~\bibnamefont{Schr{\"o}der}}, \bibinfo {author}
  {\bibfnamefont{B.~I.}\ \bibnamefont{Lundqvist}},\ and\ \bibinfo {author}
  {\bibfnamefont{D.~C.}\ \bibnamefont{Langreth}},\ }%
  \bibfield{journal}{%
  \bibinfo {journal} {J. Am. Chem. Soc.}\ }%
  \textbf{\bibinfo {volume} {130}} (\bibinfo {year} {2008})%
  \bibAnnoteFile{NoStop}{Cooper2008a}%
\bibitem{Cooper2008b}%
  \BibitemOpen
  \bibfield{author}{%
  \bibinfo {author} {\bibfnamefont{V.~R.}\ \bibnamefont{Cooper}}, \bibinfo
  {author} {\bibfnamefont{T.}~\bibnamefont{Thonhauser}},\ and\ \bibinfo
  {author} {\bibfnamefont{D.~C.}\ \bibnamefont{Langreth}},\ }%
  \bibfield{journal}{%
  \bibinfo {journal} {J. Chem. Phys.}\ }%
  \textbf{\bibinfo {volume} {128}},\ \bibinfo {pages} {204102} (\bibinfo {year}
  {2008})%
  \bibAnnoteFile{NoStop}{Cooper2008b}%
\bibitem{Thonhauser2006}%
  \BibitemOpen
  \bibfield{author}{%
  \bibinfo {author} {\bibfnamefont{T.}~\bibnamefont{Thonhauser}}, \bibinfo
  {author} {\bibfnamefont{A.}~\bibnamefont{Puzder}},\ and\ \bibinfo {author}
  {\bibfnamefont{D.~C.}\ \bibnamefont{Langreth}},\ }%
  \bibfield{journal}{%
  \bibinfo {journal} {J. Chem. Phys.}\ }%
  \textbf{\bibinfo {volume} {124}},\ \bibinfo {pages} {164106} (\bibinfo {year}
  {2006})%
  \bibAnnoteFile{NoStop}{Thonhauser2006}%
\bibitem{Hooper2008}%
  \BibitemOpen
  \bibfield{author}{%
  \bibinfo {author} {\bibfnamefont{J.}~\bibnamefont{Hooper}}, \bibinfo {author}
  {\bibfnamefont{V.~R.}\ \bibnamefont{Cooper}}, \bibinfo {author}
  {\bibfnamefont{T.}~\bibnamefont{Thonhauser}}, \bibinfo {author}
  {\bibfnamefont{N.~A.}\ \bibnamefont{Romero}}, \bibinfo {author}
  {\bibfnamefont{F.}~\bibnamefont{Zerilli}},\ and\ \bibinfo {author}
  {\bibfnamefont{D.~C.}\ \bibnamefont{Langreth}},\ }%
  \bibfield{journal}{%
  \bibinfo {journal} {ChemPhysChem}\ }%
  \textbf{\bibinfo {volume} {9}},\ \bibinfo {pages} {891} (\bibinfo {year}
  {2008})%
  \bibAnnoteFile{NoStop}{Hooper2008}%
\bibitem{Li2009}%
  \BibitemOpen
  \bibfield{author}{%
  \bibinfo {author} {\bibfnamefont{S.}~\bibnamefont{Li}}, \bibinfo {author}
  {\bibfnamefont{V.~R.}\ \bibnamefont{Cooper}}, \bibinfo {author}
  {\bibfnamefont{T.}~\bibnamefont{Thonhauser}}, \bibinfo {author}
  {\bibfnamefont{B.~I.}\ \bibnamefont{Lundqvist}},\ and\ \bibinfo {author}
  {\bibfnamefont{D.~C.}\ \bibnamefont{Langreth}},\ }%
  \bibfield{journal}{%
  \bibinfo {journal} {J. Phys. Chem. B}\ }%
  \textbf{\bibinfo {volume} {113}},\ \bibinfo {pages} {11166} (\bibinfo {year}
  {2009})%
  \bibAnnoteFile{NoStop}{Li2009}%
\bibitem{Bil2011}%
  \BibitemOpen
  \bibfield{author}{%
  \bibinfo {author} {\bibfnamefont{A.}~\bibnamefont{Bil}}, \bibinfo {author}
  {\bibfnamefont{B.}~\bibnamefont{Kolb}}, \bibinfo {author}
  {\bibfnamefont{R.}~\bibnamefont{Atkinson}}, \bibinfo {author}
  {\bibfnamefont{D.~G.}\ \bibnamefont{Pettifor}}, \bibinfo {author}
  {\bibfnamefont{T.}~\bibnamefont{Thonhauser}},\ and\ \bibinfo {author}
  {\bibfnamefont{A.~N.}\ \bibnamefont{Kolmogorov}},\ }%
  \bibfield{journal}{%
  \bibinfo {journal} {Phys. Rev. B}}%
   (\bibinfo {year} {2011}),\ \bibinfo {note} {under review}%
  \bibAnnoteFile{NoStop}{Bil2011}%
\bibitem{Hamada2010}%
  \BibitemOpen
  \bibfield{author}{%
  \bibinfo {author} {\bibfnamefont{I.}~\bibnamefont{Hamada}},\ }%
  \bibfield{journal}{%
  \bibinfo {journal} {J. Chem. Phys.}\ }%
  \textbf{\bibinfo {volume} {133}},\ \bibinfo {pages} {214503} (\bibinfo {year}
  {2010})%
  \bibAnnoteFile{NoStop}{Hamada2010}%
\bibitem{Hamada2010b}%
  \BibitemOpen
  \bibfield{author}{%
  \bibinfo {author} {\bibfnamefont{I.}~\bibnamefont{Hamada}}, \bibinfo {author}
  {\bibfnamefont{K.}~\bibnamefont{Lee}},\ and\ \bibinfo {author}
  {\bibfnamefont{Y.}~\bibnamefont{Morikawa}},\ }%
  \bibfield{journal}{%
  \bibinfo {journal} {Phys. Rev. B}\ }%
  \textbf{\bibinfo {volume} {81}} (\bibinfo {year} {2010})%
  \bibAnnoteFile{NoStop}{Hamada2010b}%
\bibitem{Klimes2010}%
  \BibitemOpen
  \bibfield{author}{%
  \bibinfo {author} {\bibfnamefont{J.}~\bibnamefont{Klimes}}, \bibinfo {author}
  {\bibfnamefont{D.~R.}\ \bibnamefont{Bowler}},\ and\ \bibinfo {author}
  {\bibfnamefont{A.}~\bibnamefont{Michaelides}},\ }%
  \bibfield{journal}{%
  \bibinfo {journal} {J. Phys.: Condensed Matter}\ }%
  \textbf{\bibinfo {volume} {22}},\ \bibinfo {pages} {022201} (\bibinfo {year}
  {2010})%
  \bibAnnoteFile{NoStop}{Klimes2010}%
\bibitem{Kelkkanen2009}%
  \BibitemOpen
  \bibfield{author}{%
  \bibinfo {author} {\bibfnamefont{A.~K.}\ \bibnamefont{Kelkkanen}}, \bibinfo
  {author} {\bibfnamefont{B.~I.}\ \bibnamefont{Lundqvist}},\ and\ \bibinfo
  {author} {\bibfnamefont{J.~K.}\ \bibnamefont{N{\o}rskov}},\ }%
  \bibfield{journal}{%
  \bibinfo {journal} {J. Chem. Phys.}\ }%
  \textbf{\bibinfo {volume} {131}},\ \bibinfo {pages} {046102} (\bibinfo {year}
  {2009})%
  \bibAnnoteFile{NoStop}{Kelkkanen2009}%
\bibitem{Wang2011}%
  \BibitemOpen
  \bibfield{author}{%
  \bibinfo {author} {\bibfnamefont{J.}~\bibnamefont{Wang}}, \bibinfo {author}
  {\bibfnamefont{G.}~\bibnamefont{Rom\'{a}n-P\'{e}rez}}, \bibinfo {author}
  {\bibfnamefont{J.~M.}\ \bibnamefont{Soler}}, \bibinfo {author}
  {\bibfnamefont{E.}~\bibnamefont{Artacho}},\ and\ \bibinfo {author}
  {\bibfnamefont{M.-V.}\ \bibnamefont{Fern\'{a}ndez-Serra}},\ }%
  \bibfield{journal}{%
  \bibinfo {journal} {J. Chem. Phys.}\ }%
  \textbf{\bibinfo {volume} {134}},\ \bibinfo {pages} {024516} (\bibinfo {year}
  {2011})%
  \bibAnnoteFile{NoStop}{Wang2011}%
\bibitem{Li2008}%
  \BibitemOpen
  \bibfield{author}{%
  \bibinfo {author} {\bibfnamefont{S.}~\bibnamefont{Li}}, \bibinfo {author}
  {\bibfnamefont{V.~R.}\ \bibnamefont{Cooper}}, \bibinfo {author}
  {\bibfnamefont{T.}~\bibnamefont{Thonhauser}}, \bibinfo {author}
  {\bibfnamefont{A.}~\bibnamefont{Puzder}}, ,\ and\ \bibinfo {author}
  {\bibfnamefont{D.~C.}\ \bibnamefont{Langreth}},\ }%
  \bibfield{journal}{%
  \bibinfo {journal} {J. Phys. Chem. A}\ }%
  \textbf{\bibinfo {volume} {112}},\ \bibinfo {pages} {9031} (\bibinfo {year}
  {2008})%
  \bibAnnoteFile{NoStop}{Li2008}%
\bibitem{QE-2009}%
  \BibitemOpen
  \bibfield{author}{%
  \bibinfo {author} {\bibfnamefont{P.}~\bibnamefont{Giannozzi}}, \bibinfo
  {author} {\bibfnamefont{S.}~\bibnamefont{Baroni}}, \bibinfo {author}
  {\bibfnamefont{N.}~\bibnamefont{Bonini}}, \bibinfo {author}
  {\bibfnamefont{M.}~\bibnamefont{Calandra}}, \bibinfo {author}
  {\bibfnamefont{R.}~\bibnamefont{Car}}, \bibinfo {author}
  {\bibfnamefont{C.}~\bibnamefont{Cavazzoni}}, \bibinfo {author}
  {\bibfnamefont{D.}~\bibnamefont{Ceresoli}}, \bibinfo {author}
  {\bibfnamefont{G.~L.}\ \bibnamefont{Chiarotti}}, \bibinfo {author}
  {\bibfnamefont{M.}~\bibnamefont{Cococcioni}}, \bibinfo {author}
  {\bibfnamefont{I.}~\bibnamefont{Dabo}}, \bibinfo {author}
  {\bibfnamefont{A.}~\bibnamefont{{Dal Corso}}}, \bibinfo {author}
  {\bibfnamefont{S.}~\bibnamefont{de~Gironcoli}}, \bibinfo {author}
  {\bibfnamefont{S.}~\bibnamefont{Fabris}}, \bibinfo {author}
  {\bibfnamefont{G.}~\bibnamefont{Fratesi}}, \bibinfo {author}
  {\bibfnamefont{R.}~\bibnamefont{Gebauer}}, \bibinfo {author}
  {\bibfnamefont{U.}~\bibnamefont{Gerstmann}}, \bibinfo {author}
  {\bibfnamefont{C.}~\bibnamefont{Gougoussis}}, \bibinfo {author}
  {\bibfnamefont{A.}~\bibnamefont{Kokalj}}, \bibinfo {author}
  {\bibfnamefont{M.}~\bibnamefont{Lazzeri}}, \bibinfo {author}
  {\bibfnamefont{L.}~\bibnamefont{Martin-Samos}}, \bibinfo {author}
  {\bibfnamefont{N.}~\bibnamefont{Marzari}}, \bibinfo {author}
  {\bibfnamefont{F.}~\bibnamefont{Mauri}}, \bibinfo {author}
  {\bibfnamefont{R.}~\bibnamefont{Mazzarello}}, \bibinfo {author}
  {\bibfnamefont{S.}~\bibnamefont{Paolini}}, \bibinfo {author}
  {\bibfnamefont{A.}~\bibnamefont{Pasquarello}}, \bibinfo {author}
  {\bibfnamefont{L.}~\bibnamefont{Paulatto}}, \bibinfo {author}
  {\bibfnamefont{C.}~\bibnamefont{Sbraccia}}, \bibinfo {author}
  {\bibfnamefont{S.}~\bibnamefont{Scandolo}}, \bibinfo {author}
  {\bibfnamefont{G.}~\bibnamefont{Sclauzero}}, \bibinfo {author}
  {\bibfnamefont{A.~P.}\ \bibnamefont{Seitsonen}}, \bibinfo {author}
  {\bibfnamefont{A.}~\bibnamefont{Smogunov}}, \bibinfo {author}
  {\bibfnamefont{P.}~\bibnamefont{Umari}},\ and\ \bibinfo {author}
  {\bibfnamefont{R.~M.}\ \bibnamefont{Wentzcovitch}},\ }%
  \bibfield{journal}{%
  \bibinfo {journal} {J. Phys.: Condensed Matter}\ }%
  \textbf{\bibinfo {volume} {21}},\ \bibinfo {pages} {395502} (\bibinfo {year}
  {2009})%
  \bibAnnoteFile{NoStop}{QE-2009}%
\bibitem{Perdew1992}%
  \BibitemOpen
  \bibfield{author}{%
  \bibinfo {author} {\bibfnamefont{J.~P.}\ \bibnamefont{Perdew}}\ and\ \bibinfo
  {author} {\bibfnamefont{Y.}~\bibnamefont{Wang}},\ }%
  \bibfield{journal}{%
  \bibinfo {journal} {Phys. Rev. B}\ }%
  \textbf{\bibinfo {volume} {45}},\ \bibinfo {pages} {13244} (\bibinfo {year}
  {1992})%
  \bibAnnoteFile{NoStop}{Perdew1992}%
\bibitem{Perdew1996}%
  \BibitemOpen
  \bibfield{author}{%
  \bibinfo {author} {\bibfnamefont{J.~P.}\ \bibnamefont{Perdew}}, \bibinfo
  {author} {\bibfnamefont{K.}~\bibnamefont{Burke}},\ and\ \bibinfo {author}
  {\bibfnamefont{M.}~\bibnamefont{Ernzerhof}},\ }%
  \bibfield{journal}{%
  \bibinfo {journal} {Phys. Rev. Lett.}\ }%
  \textbf{\bibinfo {volume} {77}},\ \bibinfo {pages} {3865} (\bibinfo {year}
  {1996})%
  \bibAnnoteFile{NoStop}{Perdew1996}%
\bibitem{Zhang1998}%
  \BibitemOpen
  \bibfield{author}{%
  \bibinfo {author} {\bibfnamefont{Y.}~\bibnamefont{Zhang}}\ and\ \bibinfo
  {author} {\bibfnamefont{W.}~\bibnamefont{Yang}},\ }%
  \bibfield{journal}{%
  \bibinfo {journal} {Phys. Rev. Lett.}\ }%
  \textbf{\bibinfo {volume} {80}},\ \bibinfo {pages} {890} (\bibinfo {year}
  {1998})%
  \bibAnnoteFile{NoStop}{Zhang1998}%
\bibitem{Gaussian}%
  \BibitemOpen
  \bibfield{author}{%
  \bibinfo {author} {\bibfnamefont{M.~J.}\ \bibnamefont{Frisch}}, \bibinfo
  {author} {\bibfnamefont{G.~W.}\ \bibnamefont{Trucks}}, \bibinfo {author}
  {\bibfnamefont{H.~B.}\ \bibnamefont{Schlegel}}, \bibinfo {author}
  {\bibfnamefont{G.~E.}\ \bibnamefont{Scuseria}}, \bibinfo {author}
  {\bibfnamefont{M.~A.}\ \bibnamefont{Rob}}, \bibinfo {author}
  {\bibfnamefont{J.~R.}\ \bibnamefont{Cheeseman}}, \bibinfo {author}
  {\bibfnamefont{J.~A.~M.}\ \bibnamefont{Jr.}}, \bibinfo {author}
  {\bibfnamefont{T.}~\bibnamefont{Vreven}}, \bibinfo {author}
  {\bibfnamefont{K.~N.}\ \bibnamefont{Kudin}}, \bibinfo {author}
  {\bibfnamefont{J.~C.}\ \bibnamefont{Burant}}, \bibinfo {author}
  {\bibfnamefont{J.~M.}\ \bibnamefont{Millam}}, \bibinfo {author}
  {\bibfnamefont{S.~S.}\ \bibnamefont{Iyengar}}, \bibinfo {author}
  {\bibfnamefont{J.}~\bibnamefont{Tomasi}}, \bibinfo {author}
  {\bibfnamefont{V.}~\bibnamefont{Barone}}, \bibinfo {author}
  {\bibfnamefont{B.}~\bibnamefont{Mennucci}}, \bibinfo {author}
  {\bibfnamefont{M.}~\bibnamefont{Cossi}}, \bibinfo {author}
  {\bibfnamefont{G.}~\bibnamefont{Scalmani}}, \bibinfo {author}
  {\bibfnamefont{N.}~\bibnamefont{Rega}}, \bibinfo {author}
  {\bibfnamefont{G.~A.}\ \bibnamefont{Petersson}}, \bibinfo {author}
  {\bibfnamefont{H.}~\bibnamefont{Nakatsuji}}, \bibinfo {author}
  {\bibfnamefont{M.}~\bibnamefont{Hada}}, \bibinfo {author}
  {\bibfnamefont{M.}~\bibnamefont{Ehara}}, \bibinfo {author}
  {\bibfnamefont{K.}~\bibnamefont{Toyota}}, \bibinfo {author}
  {\bibfnamefont{R.}~\bibnamefont{Fukuda}}, \bibinfo {author}
  {\bibfnamefont{J.}~\bibnamefont{Hasegawa}}, \bibinfo {author}
  {\bibfnamefont{M.}~\bibnamefont{Ishida}}, \bibinfo {author}
  {\bibfnamefont{T.}~\bibnamefont{Nakajima}}, \bibinfo {author}
  {\bibfnamefont{Y.}~\bibnamefont{Honda}}, \bibinfo {author}
  {\bibfnamefont{O.}~\bibnamefont{Kitao}}, \bibinfo {author}
  {\bibfnamefont{H.}~\bibnamefont{Nakai}}, \bibinfo {author}
  {\bibfnamefont{M.}~\bibnamefont{Klene}}, \bibinfo {author}
  {\bibfnamefont{X.}~\bibnamefont{Li}}, \bibinfo {author}
  {\bibfnamefont{J.~E.}\ \bibnamefont{Knox}}, \bibinfo {author}
  {\bibfnamefont{H.~P.}\ \bibnamefont{Hratchian}}, \bibinfo {author}
  {\bibfnamefont{J.~B.}\ \bibnamefont{Cross}}, \bibinfo {author}
  {\bibfnamefont{V.}~\bibnamefont{Bakken}}, \bibinfo {author}
  {\bibfnamefont{C.}~\bibnamefont{Adamo}}, \bibinfo {author}
  {\bibfnamefont{J.}~\bibnamefont{Jaramillo}}, \bibinfo {author}
  {\bibfnamefont{R.}~\bibnamefont{Gomperts}}, \bibinfo {author}
  {\bibfnamefont{R.~E.}\ \bibnamefont{Stratmann}}, \bibinfo {author}
  {\bibfnamefont{O.}~\bibnamefont{Yazyev}}, \bibinfo {author}
  {\bibfnamefont{A.~J.}\ \bibnamefont{Austin}}, \bibinfo {author}
  {\bibfnamefont{R.}~\bibnamefont{Cammi}}, \bibinfo {author}
  {\bibfnamefont{C.}~\bibnamefont{Pomelli}}, \bibinfo {author}
  {\bibfnamefont{J.~W.}\ \bibnamefont{Ochterski}}, \bibinfo {author}
  {\bibfnamefont{P.~Y.}\ \bibnamefont{Ayala}}, \bibinfo {author}
  {\bibfnamefont{K.}~\bibnamefont{Morokuma}}, \bibinfo {author}
  {\bibfnamefont{G.~A.}\ \bibnamefont{Voth}}, \bibinfo {author}
  {\bibfnamefont{P.}~\bibnamefont{Salvador}}, \bibinfo {author}
  {\bibfnamefont{J.~J.}\ \bibnamefont{Dannenberg}}, \bibinfo {author}
  {\bibfnamefont{V.~G.}\ \bibnamefont{Zakrzewski}}, \bibinfo {author}
  {\bibfnamefont{S.}~\bibnamefont{Dapprich}}, \bibinfo {author}
  {\bibfnamefont{A.~D.}\ \bibnamefont{Daniels}}, \bibinfo {author}
  {\bibfnamefont{M.~C.}\ \bibnamefont{Strain}}, \bibinfo {author}
  {\bibfnamefont{O.}~\bibnamefont{Farkas}}, \bibinfo {author}
  {\bibfnamefont{D.~K.}\ \bibnamefont{Malick}}, \bibinfo {author}
  {\bibfnamefont{A.~D.}\ \bibnamefont{Rabuck}}, \bibinfo {author}
  {\bibfnamefont{K.}~\bibnamefont{Raghavachari}}, \bibinfo {author}
  {\bibfnamefont{J.~B.}\ \bibnamefont{Foresman}}, \bibinfo {author}
  {\bibfnamefont{J.~V.}\ \bibnamefont{Ortiz}}, \bibinfo {author}
  {\bibfnamefont{Q.}~\bibnamefont{Cui}}, \bibinfo {author}
  {\bibfnamefont{A.~G.}\ \bibnamefont{Baboul}}, \bibinfo {author}
  {\bibfnamefont{S.}~\bibnamefont{Clifford}}, \bibinfo {author}
  {\bibfnamefont{J.}~\bibnamefont{Cioslowski}}, \bibinfo {author}
  {\bibfnamefont{B.~B.}\ \bibnamefont{Stefanov}}, \bibinfo {author}
  {\bibfnamefont{G.}~\bibnamefont{Liu}}, \bibinfo {author}
  {\bibfnamefont{A.}~\bibnamefont{Liashenko}}, \bibinfo {author}
  {\bibfnamefont{P.}~\bibnamefont{Piskorz}}, \bibinfo {author}
  {\bibfnamefont{I.}~\bibnamefont{Komaromi}}, \bibinfo {author}
  {\bibfnamefont{R.~L.}\ \bibnamefont{Martin}}, \bibinfo {author}
  {\bibfnamefont{D.~J.}\ \bibnamefont{Fox}}, \bibinfo {author}
  {\bibfnamefont{T.}~\bibnamefont{Keith}}, \bibinfo {author}
  {\bibfnamefont{M.~A.}\ \bibnamefont{Al-Laham}}, \bibinfo {author}
  {\bibfnamefont{C.~Y.}\ \bibnamefont{Peng}}, \bibinfo {author}
  {\bibfnamefont{A.}~\bibnamefont{Nanayakkara}}, \bibinfo {author}
  {\bibfnamefont{M.}~\bibnamefont{Challacombe}}, \bibinfo {author}
  {\bibfnamefont{P.~M.~W.}\ \bibnamefont{Gill}}, \bibinfo {author}
  {\bibfnamefont{B.}~\bibnamefont{Johnson}}, \bibinfo {author}
  {\bibfnamefont{W.}~\bibnamefont{Chen}}, \bibinfo {author}
  {\bibfnamefont{M.~W.}\ \bibnamefont{Wong}}, \bibinfo {author}
  {\bibfnamefont{C.}~\bibnamefont{Gonzalez}},\ and\ \bibinfo {author}
  {\bibfnamefont{J.~A.}\ \bibnamefont{Pople}},\ }%
  \enquote{\bibinfo {title} {Gaussian~03 {R}evision d.02},}\ \bibinfo {note}
  {Gaussian, Inc., Wallingford CT, 2004.}%
  \bibAnnoteFile{Stop}{Gaussian}%
\bibitem{Santra2007}%
  \BibitemOpen
  \bibfield{author}{%
  \bibinfo {author} {\bibfnamefont{B.}~\bibnamefont{Santra}}, \bibinfo {author}
  {\bibfnamefont{A.}~\bibnamefont{Michaelides}},\ and\ \bibinfo {author}
  {\bibfnamefont{M.}~\bibnamefont{Scheffler}},\ }%
  \bibfield{journal}{%
  \bibinfo {journal} {J. Chem. Phys.}\ }%
  \textbf{\bibinfo {volume} {127}} (\bibinfo {year} {2007})%
  \bibAnnoteFile{NoStop}{Santra2007}%
\bibitem{Petrenko1999}%
  \BibitemOpen
  \bibfield{author}{%
  \bibinfo {author} {\bibfnamefont{V.~F.}\ \bibnamefont{Petrenko}}\ and\
  \bibinfo {author} {\bibfnamefont{R.~W.}\ \bibnamefont{Whitworth}},\ }%
  \emph{\bibinfo {title} {{Physics of ice}}}\ (\bibinfo {publisher} {Oxford
  University Press, USA},\ \bibinfo {year} {1999})\ p.\ \bibinfo {pages} {390}%
  \bibAnnoteFile{NoStop}{Petrenko1999}%
\bibitem{Liu1996}%
  \BibitemOpen
  \bibfield{author}{%
  \bibinfo {author} {\bibfnamefont{K.}~\bibnamefont{Liu}}, \bibinfo {author}
  {\bibfnamefont{J.~D.}\ \bibnamefont{Cruzan}},\ and\ \bibinfo {author}
  {\bibfnamefont{R.~J.}\ \bibnamefont{Saykally}},\ }%
  \bibfield{journal}{%
  \bibinfo {journal} {Science}\ }%
  \textbf{\bibinfo {volume} {271}},\ \bibinfo {pages} {929} (\bibinfo {year}
  {1996})%
  \bibAnnoteFile{NoStop}{Liu1996}%
\bibitem{Bergmann2007}%
  \BibitemOpen
  \bibfield{author}{%
  \bibinfo {author} {\bibfnamefont{U.}~\bibnamefont{Bergmann}}, \bibinfo
  {author} {\bibfnamefont{A.}~\bibnamefont{{Di Cicco}}}, \bibinfo {author}
  {\bibfnamefont{P.}~\bibnamefont{Wernet}}, \bibinfo {author}
  {\bibfnamefont{E.}~\bibnamefont{Principi}}, \bibinfo {author}
  {\bibfnamefont{P.}~\bibnamefont{Glatzel}},\ and\ \bibinfo {author}
  {\bibfnamefont{A.}~\bibnamefont{Nilsson}},\ }%
  \bibfield{journal}{%
  \bibinfo {journal} {J. Chem. Phys.}\ }%
  \textbf{\bibinfo {volume} {127}},\ \bibinfo {pages} {174504} (\bibinfo {year}
  {2007})%
  \bibAnnoteFile{NoStop}{Bergmann2007}%
\bibitem{Gregory1998}%
  \BibitemOpen
  \bibfield{author}{%
  \bibinfo {author} {\bibfnamefont{J.~K.}\ \bibnamefont{Gregory}},\ }%
  \bibfield{journal}{%
  \bibinfo {journal} {Chem. Phys. Lett.}\ }%
  \textbf{\bibinfo {volume} {282}},\ \bibinfo {pages} {147} (\bibinfo {year}
  {1998})%
  \bibAnnoteFile{NoStop}{Gregory1998}%
\bibitem{Ceponkus2008}%
  \BibitemOpen
  \bibfield{author}{%
  \bibinfo {author} {\bibfnamefont{J.}~\bibnamefont{Ceponkus}}, \bibinfo
  {author} {\bibfnamefont{P.}~\bibnamefont{Uvdal}},\ and\ \bibinfo {author}
  {\bibfnamefont{B.}~\bibnamefont{Nelander}},\ }%
  \bibfield{journal}{%
  \bibinfo {journal} {J. Chem. Phys.}\ }%
  \textbf{\bibinfo {volume} {129}},\ \bibinfo {pages} {194306} (\bibinfo {year}
  {2008})%
  \bibAnnoteFile{NoStop}{Ceponkus2008}%
\bibitem{Tremblay2010}%
  \BibitemOpen
  \bibfield{author}{%
  \bibinfo {author} {\bibfnamefont{B.}~\bibnamefont{Tremblay}}, \bibinfo
  {author} {\bibfnamefont{B.}~\bibnamefont{Madeb\`{e}ne}}, \bibinfo {author}
  {\bibfnamefont{M.~E.}\ \bibnamefont{Alikhani}},\ and\ \bibinfo {author}
  {\bibfnamefont{J.~P.}\ \bibnamefont{Perchard}},\ }%
  \bibfield{journal}{%
  \bibinfo {journal} {Chem. Phys.}\ }%
  \textbf{\bibinfo {volume} {378}},\ \bibinfo {pages} {27} (\bibinfo {year}
  {2010})%
  \bibAnnoteFile{NoStop}{Tremblay2010}%
\bibitem{Zhang2011}%
  \BibitemOpen
  \bibfield{author}{%
  \bibinfo {author} {\bibfnamefont{C.}~\bibnamefont{Zhang}}, \bibinfo {author}
  {\bibfnamefont{D.}~\bibnamefont{Donadio}}, \bibinfo {author}
  {\bibfnamefont{F.}~\bibnamefont{Gygi}},\ and\ \bibinfo {author}
  {\bibfnamefont{G.}~\bibnamefont{Galli}},\ }%
  \bibfield{journal}{%
  \bibinfo {journal} {J. Chem. Theory Comput.}\ }%
  \textbf{\bibinfo {volume} {7}},\ \bibinfo {pages} {1443} (\bibinfo {year}
  {2011})%
  \bibAnnoteFile{NoStop}{Zhang2011}%
\bibitem{Roman-Perez2009}%
  \BibitemOpen
  \bibfield{author}{%
  \bibinfo {author} {\bibfnamefont{G.}~\bibnamefont{Rom\'{a}n-P\'{e}rez}}\ and\
  \bibinfo {author} {\bibfnamefont{J.~M.}\ \bibnamefont{Soler}},\ }%
  \bibfield{journal}{%
  \bibinfo {journal} {Phys. Rev. Lett.}\ }%
  \textbf{\bibinfo {volume} {103}},\ \bibinfo {pages} {096102} (\bibinfo {year}
  {2009})%
  \bibAnnoteFile{NoStop}{Roman-Perez2009}%
\bibitem{Lee2010}%
  \BibitemOpen
  \bibfield{author}{%
  \bibinfo {author} {\bibfnamefont{K.}~\bibnamefont{Lee}}, \bibinfo {author}
  {\bibfnamefont{E.~D.}\ \bibnamefont{Murray}}, \bibinfo {author}
  {\bibfnamefont{L.}~\bibnamefont{Kong}}, \bibinfo {author}
  {\bibfnamefont{B.~I.}\ \bibnamefont{Lundqvist}},\ and\ \bibinfo {author}
  {\bibfnamefont{D.~C.}\ \bibnamefont{Langreth}},\ }%
  \bibfield{journal}{%
  \bibinfo {journal} {Phys. Rev. B}\ }%
  \textbf{\bibinfo {volume} {82}},\ \bibinfo {pages} {081101(R)} (\bibinfo
  {year} {2010})%
  \bibAnnoteFile{NoStop}{Lee2010}%
\end{thebibliography}
\end{document}